\documentclass[referee,a4paper,12pt,traditabstract]{swsc} 

\usepackage{graphicx}
\usepackage{subfigure}
\usepackage{epstopdf}
\usepackage[displaymath,mathlines]{lineno}
\usepackage[authoryear,round]{natbib}
\usepackage[backref]{hyperref}
\usepackage{url}

\hypersetup{colorlinks=true,citecolor=cyan,urlcolor=cyan,linkcolor=blue}

\graphicspath{{figs2/},{./}}
\def\degdot{\hbox{$.\!\!^\circ$}}
\newcommand{\ardor}{\mbox{\sl ARDoR}}
\newcommand{\fix}[1]{{#1}} 

\begin{document}

%\begin{linenumbers}  

\title{Towards an algebraic method of solar cycle prediction}
\subtitle{II. Reducing the need for detailed input data with $\ardor$}

\titlerunning{Algebraic method of solar cycle prediction II.}
\authorrunning{Nagy et al.}

\author{Melinda Nagy\inst{1}
\and 
Krist\'of Petrovay\inst{1} 
\and 
Alexandre Lemerle\inst{2,3}
\and 
Paul Charbonneau\inst{2}
}

\institute{E\"otv\"os Lor\'and University, Department of Astronomy, Budapest,
Hungary\\
\email{\href{mailto:M.Nagy@astro.elte.hu}{M.Nagy@astro.elte.hu},
\href{mailto:K.Petrovay@astro.elte.hu}{K.Petrovay@astro.elte.hu}}
\and
D\'epartement de Physique, Universit\'e de Montr\'eal, Montr\'eal, QC, Canada\\
\email{\href{mailto:paulchar@astro.umontreal.ca}{paulchar@astro.umontreal.ca},
\href{mailto:lemerle@astro.umontreal.ca}{lemerle@astro.umontreal.ca}}
\and
{Bois-de-Boulogne College, Montr\'eal, QC, Canada}
}

%%   \date{Received September 15, 1996; accepted March 16, 1997}

  % \abstract{}{}{}{}{}        %% uncomment if structured abstract is desired
 %% 5 {} token are mandatory
 
\abstract
%% context heading (optional). leave {} empty if necessary  {}
%% aims heading (mandatory)  {}
%% methods heading (mandatory)  {}
%% results heading (mandatory)  {}
%% conclusions heading (optional), leave {} empty if necessary  {} 
{         
An algebraic method for the reconstruction and potentially prediction
of the solar dipole moment value at sunspot minimum (known to be a
good predictor of the amplitude of the next solar cycle) was suggested
in the first paper in this series. The method sums up the ultimate
dipole moment contributions of individual active regions in a solar
cycle: for this, detailed and reliable input data would in principle
be needed for thousands of active regions in a solar cycle. To reduce
the need for detailed input data, here we propose a new active region
descriptor called $\ardor$ (Active Region Degree of Rogueness). In a
detailed statistical analysis of a large number of activity cycles
simulated with the 2$\times$2D dynamo model we demonstrate that
ranking active regions by decreasing $\ardor$, for a good reproduction of
the solar dipole moment at the end of the cycle it is sufficient to
consider the top $N$ regions on this list explicitly, where $N$ is a
relatively low number, while for the other regions the $\ardor$ value may
be set to zero. E.g., with $N=5$  the fraction of cycles where the
dipole moment is reproduced with an error exceeding {$\pm 30$\,\% 
is only 12\%, significantly reduced with respect to the case $N=0$,
i.e.\ $\ardor$ set to zero for all active regions, where this fraction is
26\%}.  This indicates that stochastic effects on the intercycle
variations of solar activity are dominated by the effect of a low
number of large ``rogue'' active regions, rather than the combined
effect of numerous small ARs. The method has a potential for future
use in solar cycle prediction.
}        

   %% replace by pair of curly brackets, {}, if structured abstract is selected


   \maketitle
%%
%%________________________________________________________________

\section{General introduction}

The magnetic fields responsible for solar activity phenomena emerge
into the solar atmosphere in a concentrated form, in active regions
(ARs). In each solar cycle thousands of active regions are listed in
the official NOAA database and many more small active regions are
missed if their heliographic position and lifetime do not render them
directly observable on the visible hemisphere. Similarly, no detailed
catalogues exist for the ubiquitous ephemeral active regions, of even
smaller size. 

The emergence of this large number of (typically) bipolar magnetic
regions obeys some well known statistical regularities like Hale's
polarity rules and Joy's law. As a consequence, upon their decay by
turbulent diffusion their remains contribute to the large-scale
ordered photospheric magnetic field, including the Sun's global axial
dipole field {(the so-called Babcock--Leighton mechanism)}. 
Active regions emerging in a given solar cycle
contribute on average to the global dipole with a sign opposite to the
preexisting field at the start of the cycle, and these contributions
from active regions add up until, some time in the middle of the
cycle, the global field reverses and a new cycle starts at the Sun's
poles, still overlapping with the ongoing cycle at low latitudes. Flux
emergence is thus an important element of the solar dynamo mechanism
sustaining the periodically overturning solar magnetic field.

The inherently stochastic nature of flux emergence introduces random
fluctuations into this statistically ordered process. In recent years
it has been realized that the random nature of flux emergence can give
rise to significant deviations of the solar dipole moment built up
during a cycle from its expected mean value: in some cycles a small
number of so-called ``rogue'' active regions (\citealt{Petrovay:IAUS340})
with atypical properties may lead to a major, unexpected change in the
level of activity. The unexpected change in the level of activity from
solar cycle 23 to 24 has been interpreted as the result a few such
abnormal regions by \cite{Jiang+:cyc24hindcast}, while in a dynamo model
\cite{Nagy+:rogue} found that in extreme cases even a single rogue AR
can trigger a grand minimum.

An open question is how to identify the [candidate] rogue active
regions, and how many such regions need to be considered in individual
detail in models aiming to reproduce the evolution of the Sun's large
scale field. It is not a priori clear that this number is low, so the
question we pose in this paper is whether the stochastic effects in
cycle-to-cycle variation originating in the random nature of the flux
emergence process are dominated by a few ``rogue'' AR in each cycle
with individually large and unusual contributions to the dipole
moment, or by the ``fluctuation background'' due to numerous other AR
with individually much lower deviations from the expected dipole
contribution. While the recent studies cited above stressed the
importance of a few large rogue AR, the importance of the fluctuation
background cannot be discarded out of hand. The issue has obvious
practical significance from the point of view of solar cycle
prediction: it would be useful to know how many (and exactly which)
observed individual AR need to be assimilated into a model for
successful forecasts.

A related investigation was recently carried out by
\cite{Whitbread+:dipcontr}. In that work ARs were ordered by their
individual contributions to the global axial dipole moment: it was
found that, far from being dominated by a few ARs with the largest
contributions, the global dipole moment built up during a cycle cannot
be reproduced without taking into account a large number (hundreds) of
ARs. In another recent work \cite{Cameron+:toroidal} found that even
ephemeral active regions contribute to the net toroidal flux loss from
the Sun by an amount comparable to the contribution of large active
regions. By analogy, this opens the possibility that ephemeral ARs may
also contribute to the global poloidal field by a non-negligible amount,
though statistical studies of the orientation of ephemeral ARs are
unfortunately rare (cf.~\citealt{Tlatov_:ER}).

While these interesting results shed new light on the overall role of
flux emergence in smaller bipoles in the global dynamo, we think that
from the point of view of solar cycle prediction, instead of the
dipole moment contribution per se, a more relevant control parameter
is the {\it deviation} of the dipole contribution from the case with
no random fluctations in flux emergence, i.e. the ``degree of
rogueness'' (DoR). We therefore set out to systematically study the
effect of individual AR on the subsequent course of solar activity
using the DoR as an ordering parameter.

The question immediately arises how this DoR should be defined. 

The approach we take in this work assumes that the effect of random
fluctuations manifests itself primarily in the properties of
individual active regions, rather than in their spatiotemporal
distribution. The DoR based on individual AR properties will be called
``active region degree of rogueness'' --- $\ardor$ for brevity.

The structure of this paper is as follows. Section 2 introduces and 
discusses our definition of $\ardor$. In Section 3, after recalling
salient features of the 2$\times$2D dynamo model, we use statistics based
on this model to answer the central question of this paper.
Conclusions are drawn in Section 4.

\section{Introducing $\ardor$}

The Sun's axial dipolar moment is expressed as
\begin{equation}                       
    D(t) = \frac32 \int_{-\pi/2}^{\pi/2} 
    B(\lambda,t)\sin\lambda\cos\lambda\, \mathrm{d}\lambda ,
 \label{eq:dipmom}
\end{equation}
where $B$ is the azimuthal average of the large scale photospheric
magnetic field (assumed to be radial) while $\lambda$ is heliographic
latitude.

The value $D_n$ of this dipole moment at the start of cycle $n$ is
widely considered the best physics-based precursor of the the
amplitude of the incipient cyle $n$ (\citealt{Petrovay:LRSP2}). 
Understanding intercycle variations in solar activity and potentially
extending the scope of the prediction calls for an effective and
robust method to compute $D_n$ from (often limited) observational data
on the previous course of solar activity.

In the first paper of this series, \cite{Petrovay+:algebraic1}
(hereafter Paper 1) we suggested a simplified approach to the
computation of the evolution of the global axial dipole moment of the
Sun. Instead of solving the partial differential equation normally
used for modeling surface magnetic flux transport (SFT) processes on
the Sun, this method simply represents the dipole moment by an
algebraic sum: 
\begin{equation}
\Delta D_n\equiv D_{n+1} - D_n = 
  \sum_{i=1}^{N_{\mathrm{tot}}} \delta D_{U,i} =
  \sum_{i=1}^{N_{\mathrm{tot}}} \delta D_{\infty,i} \, e^{\fix{(t_i-t_{n+1})}/\tau} =
  \sum_{i=1}^{N_{\mathrm{tot}}}  
  {f_{\infty,i}}\, \delta D_{1,i} \, e^{\fix{(t_i-t_{n+1})}/\tau} ,
\label{eq:cycledipmom}
\end{equation}
where $i$ indexes the active regions in a cycle, $N_\mathrm{tot}$ is
the total number of ARs in the cycle, $\delta D_1$ is the {\it
initial} contribution of an active region to the global dipole moment,
$\delta D_U$ is its {\it ultimate} contribution at the end of a cycle
and $\tau\le\infty$ is the assumed timescale of magnetic field decay
due to radial diffusion. Furthermore, 
\begin{equation}
f_\infty=\delta D_\infty/\delta D_1 ,
\end{equation}
where $\delta D_\infty$ is the {\it asymptotic} contribution of the
same AR in a SFT model with $\tau=\infty$, once the meridional flow
has concentrated the relic magnetic flux from the AR to two opposite
polarity patches at the two poles. (See Paper 1 for further
explanations.)

In this approach, ARs are assumed to be represented by simple bipoles
at the time of their introduction into the model so their initial dipole
moment contribution is given by
\begin{equation}
\delta D_1=\frac 3{4\pi R^2}\, \Phi\, d_\lambda\cos\lambda_0 ,
\end{equation}
where $\Phi$ is the magnetic flux in the northern polarity patch, 
$d_\lambda$ is the latitudinal separation of the
polarities\footnote{$d_\lambda=d\sin\alpha$ where $d$ is the full
angular polarity separation on the solar surface and $\alpha$ is the
tilt angle of the bipole axis relative to the east--west direction,
the sign of $\alpha$ being negative for bipoles disobeying Hale's
polarity rules.},
$\lambda_0$ is the initial latitude of [the center of] the bipole and
$R$ is the radius of the Sun. As demonstrated in Paper 1, $f_\infty$,
is in turn given by
\begin{equation}
f_\infty= \frac a{\lambda_R} \exp
\left(\frac{-\lambda_0^2}{2\lambda_R^2}\right) .
\end{equation}
It was numerically demonstrated in Paper 1 that this Gaussian form
holds quite generally irrespective of the details of the SFT model, 
its parameters ($\lambda_R$ and $a$) only have a very weak dependence
on the assumed form of the meridional flow profile (at least for
profiles that are closer to observations), and their value only
depends on a single combination of SFT model parameters. The values of
$\lambda_R$ and $a$ for a given SFT model may be determined by
interpolation of the numerical results, as
presented in Paper 1.

The terms of the sum (\ref{eq:cycledipmom}) represent the ultimate
dipole contributions $\delta D_U$ of individual active regions in a
cycle at the solar minimum ending that cycle. In principle each and
every active region should be represented by an explicit term in the
sum. Such a case was indeed considered in Paper 1 in a comparison with
a run result from the 2$\times$2D dynamo and it was found that the
algebraic method returns the total change in dipole moment during a
cycle quite accurately in the overwhelming majority of cycles.

When it comes to applying the method to the real Sun, however, the
need to include each bipolar region in the source becomes quite a
nuisance. As discussed above in the Introduction, data for individual
active regions are often missing for the smaller ARs, while in the
case of the larger, more complex AR representing them by an
instantaneously introduced bipole is nontrivial. As it was recently
pointed out by \cite{Iijima+:asym}, for an AR with zero tilt but
different extents of the two polarity distributions $\delta D_\infty$
will be nonzero, even though $\delta D_1=0$ for this configuration.
The reason is that the configuration has a nonzero quadrupole moment,
which may alternatively be represented by not one but two oppositely
oriented dipoles slightly shifted in latitude.

Such intricacies would certainly make it advisable to keep the number
of active regions explicitly represented in the sum
(\ref{eq:cycledipmom}) to a minimum. This again brings us to the
central question of this paper: how many and which active regions need
to be explicitly taken into consideration for the calculation of the
solar dipole moment? While the previous study of
\cite{Whitbread+:dipcontr} has shown that keeping only a few ARs in
the summation is certainly not correct, representing the rest of the
ARs in a less faithful or detailed manner may still be admissible as
long as this does not distort the statistics. 

To select those few ARs that still need to be realistically
represented \fix{we introduce the concept of $\ardor$}. %active region degree of rogueness ($\ardor$)
As known examples of rogue AR presented e.g. in
Nagy et al. (2017) are primarily rogue on account of their unusual
tilts and large separations, the first idea is to define $\ardor$ as
the difference between the ultimate dipole moment contribution of an
AR and the value this would take with no scatter in the tilt and
separation (i.e. if the tilt and separation were to take their
expected values for the given latitude and magnetic flux,
as given by eqs.~(15) and (16a) in \citealt{Lemerle1}).

In the present paper we thus consider the case where for the majority
of ARs only the information regarding their size (magnetic flux) and
heliographic latitude is retained, while further details such as
polarity separation or tilt angle (and therefore $\delta D_1$) are
simply set to their expected values for the ARs with the given flux
and heliographic latitude (``reduced stochasticity'' or RS
representation), and compare this with the case when the actual
polarity separations and tilts are used (``fully stochastic'' or FS
case). 
The active region degree of rogueness is defined by
\begin{equation}
\ardor = \delta D_{U,\mathrm{FS}}-\delta D_{U,\mathrm{RS}} = 
f_{\infty}\, e^{(t_i-t_{n+1})/\tau} 
(\delta D_{1,\mathrm{FS}}-\delta D_{1,\mathrm{RS}}) .
\end{equation}

An objection to this definition may be raised as a large AR with
unusually low separation and/or tilt will yield a negligible
contribution to the dipole moment ($\delta D_U=0$), yet it may be
characterized by a large negative DoR value according to the proposed
definition. On the other hand, this is arguably not a shortcoming of
the approach: on the contrary, as the total flux emerging in a cycle
of a given amplitude is more or less fixed, the emergence of a large
AR with unusually low $\delta D_U$ implies that the expected 
$\delta D_U$ contribution will be ``missing'' at the final
account, resulting in the buildup of lower-than-expected global dipole
moment at the end of the cycle.

Ranking the ARs in a cycle according to their decreasing $\ardor$
values, we now set out to compare the results where $\ardor$ is explicitly
considered for the top $N$ ARs on this list, while the rest of the ARs
are represented in the RS approach. We ask the question what is the
lowest value for $N$ for which the algebraic method still yields
acceptable results?

\begin{figure}
\includegraphics[width=0.5\textwidth]{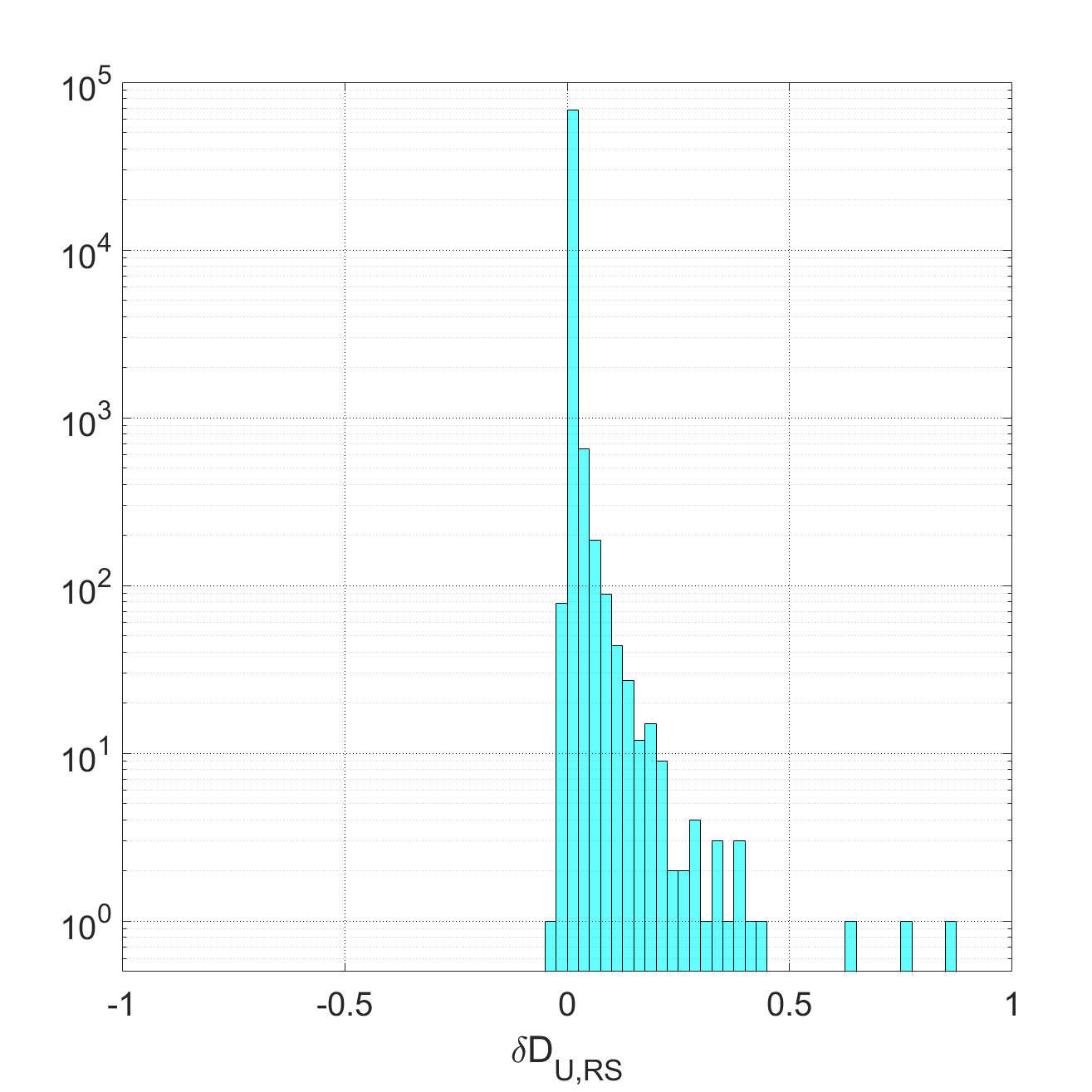}
\includegraphics[width=0.5\textwidth]{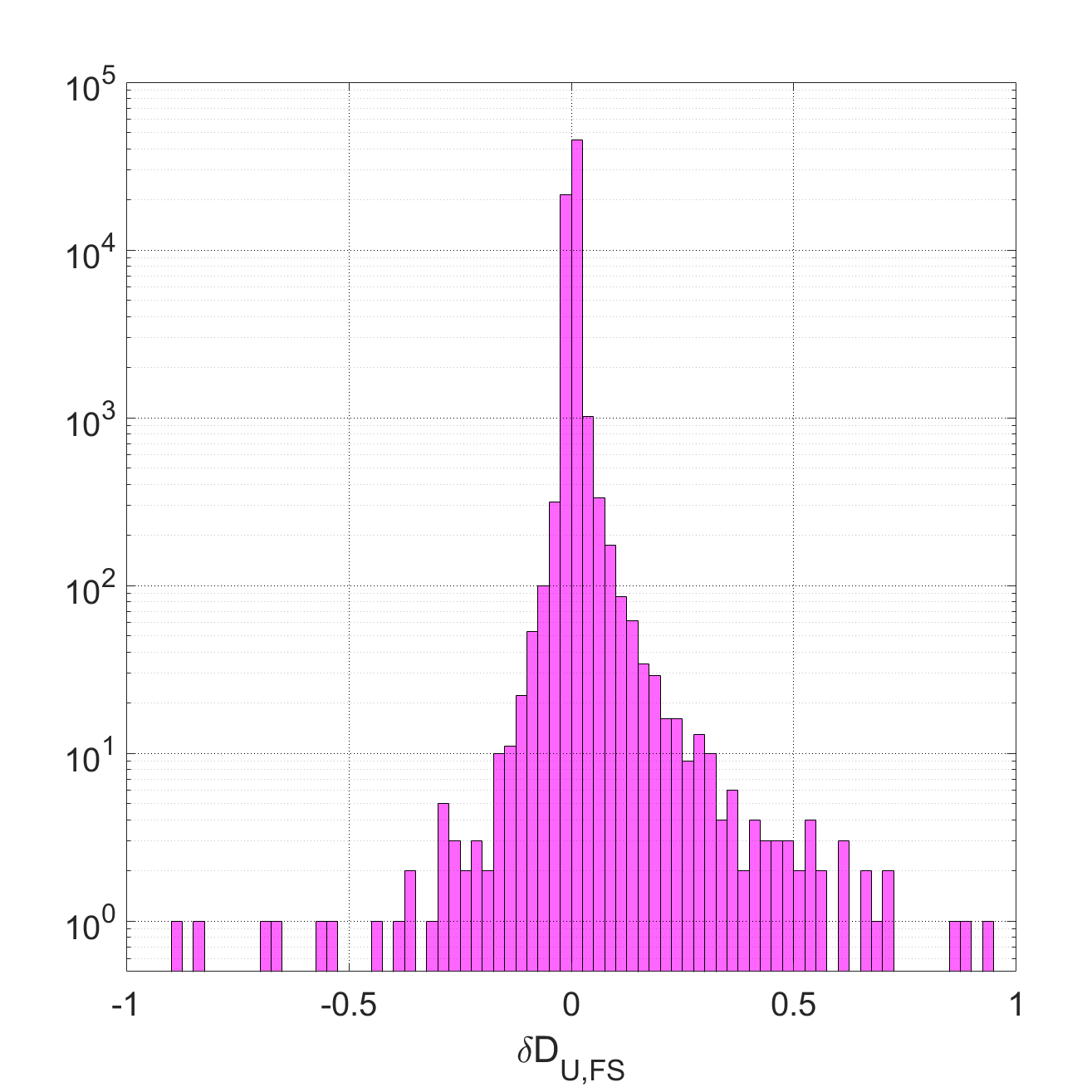}
\includegraphics[width=0.5\textwidth]{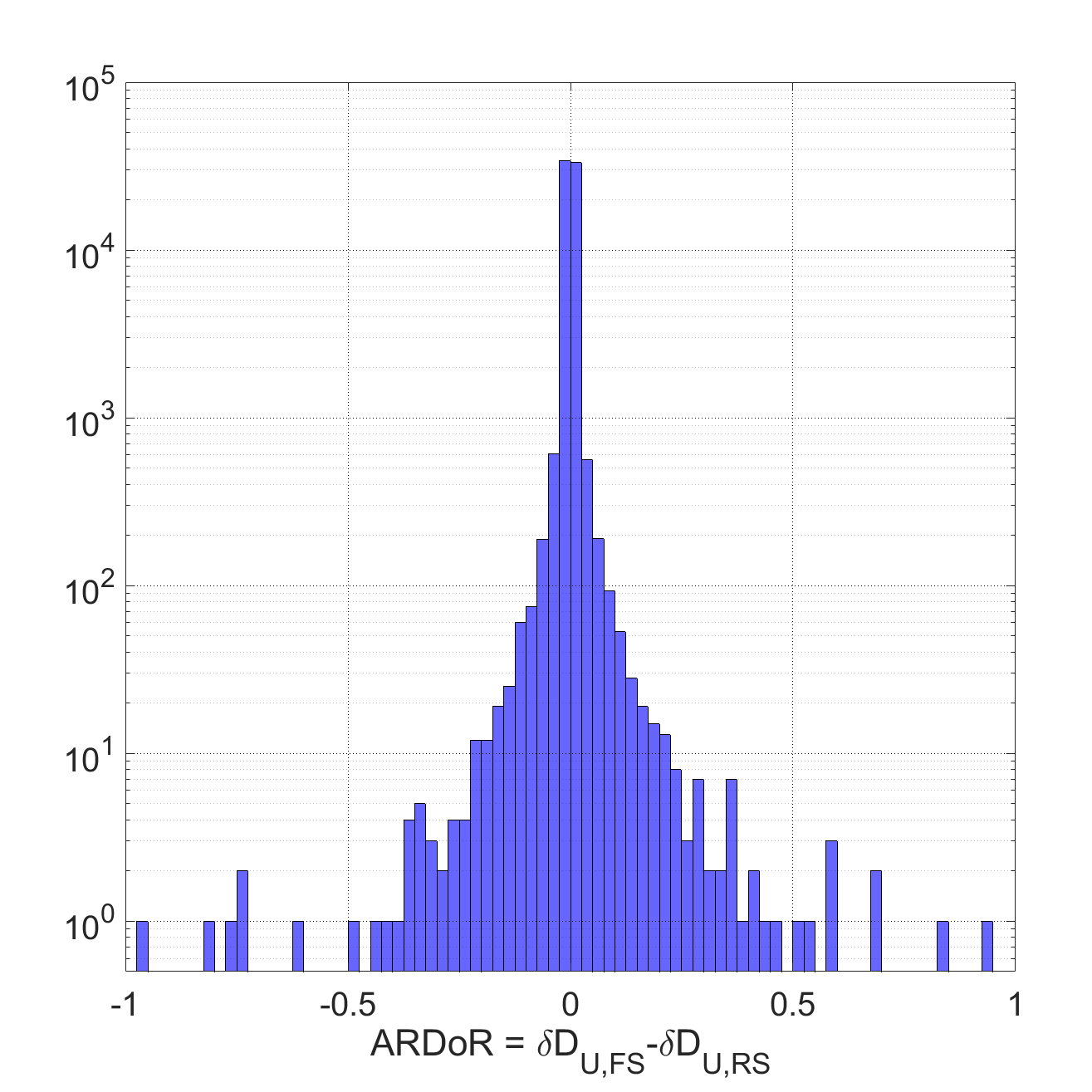}
\caption{Histograms of ultimate dipole moment contributions of
individual active regions in the FS and RS cases and their differences
(i.e. $\ardor$ values) \fix{measured in {Gauss}},  based on 647 
cycles with an average of 3073 active regions per cycle.}
\label{fig:ARDORhist}
\end{figure}

\begin{table}
\centering
\begin{tabular}{r c c c}
\hline
 N & mean & median & st.dev.\\
\hline
  1 & 0.4977 & 0.4565 & 0.4146 \\
  2 & 0.6184 & 0.6022 & 0.4305 \\
  3 & 0.6696 & 0.6735 & 0.4065 \\
  4 & 0.6996 & 0.7188 & 0.3953 \\
  5 & 0.7245 & 0.7490 & 0.3535 \\
 10 & 0.8139 & 0.8078 & 0.3136 \\
 20 & 0.8838 & 0.8822 & 0.2576 \\
 50 & 0.9381 & 0.9472 & 0.1917 \\
100 & 0.9689 & 0.9816 & 0.1324 \\
\hline\\
\end{tabular}
\caption{Means, medians and standard deviations of the total ARDoR of
the top $N$ ARs divided by the total ARDoR of all ARs for cycles
where the total ARDoR exceeds 15\,\% of the absolute change in the
dipole moment $\Delta D$ (230 cycles)}
\label{table:ARDORhist15}
\end{table}

\section{$\ardor$ and rogue active regions in the 2$\times$2D dynamo model}

Characteristics of the hybrid kinematic $2\times2$D Babcock-Leighton 
dynamo model developed by \cite{Lemerle1} and \cite{Lemerle2} are
particularly suitable for a study of this type. This model couples an
internal axially symmetric flux transport dynamo (FTD) with a surface
flux transport (SFT) model. The FTD component module provides the new
active region emergences acting as a source term in the SFT component,
while the output of the SFT model is used as upper boundary condition
on the FTD model. In the model, bipolar magnetic regions (BMRs)
representing active regions are generated at the surface randomly, 
with a probability based on the amplitude of the toroidal field
in the deep convective zone, their properties being drawn from a
statistical ensemble constructed to obey observationally determined
statistical relationships. This makes it straightforward to extract
the set of AR properties for any cycle from the model and to convert it
to a reduced stochasticity set by setting the random fluctuations
around the mean in the  distributions of polarity separations and
tilts to zero. In addition, the numerical efficiency of the model
allows to run it for a large number of simulated solar cycles,
rendering it suitable for statistical analysis of the results.

For the present analysis we use run results from the standard setup of
the 2$\times$2D model as described in \cite{Lemerle2}.  Evaluating the
parameters of the algebraic model from the numerical results presented
in Paper 1 (for the same meridional flow and parameter values as in the
dynamo model) yields $\lambda_R=13{\degdot}6$ and $a/\lambda_R=3.75$,
so for the algebraic method these values are used. The number of
simulated cycles used in the analysis was 647. The distribution of
computed $\ardor$ values is plotted in Fig.~\ref{fig:ARDORhist}.

ARs in each cycle are ranked by the $\ardor$ values. In each cyle we
compute the absolute change $\Delta D$ in the global solar dipole
moment from equation (\ref{eq:cycledipmom}) for a ``cocktail'' of ARs,
taking the ARs with the top $N$ highest $\ardor$ from the original,
fully stochastic set, while taking the rest from the RS set. For
brevity, this will be referred to as the ``rank-$N$ $\ardor$ method''.
The dipole moment change calculated with the rank-$N$ $\ardor$ method
is then
\begin{equation}
\Delta D_{\ardor,N}=\Delta D_{\mathrm{RS}} +\sum_{i=1}^N \ardor_i
\end{equation}
where the AR index $i$ is in the order of decreasing $\ardor$.

\begin{figure}
\begin{center}
\includegraphics[width=0.4\textwidth]{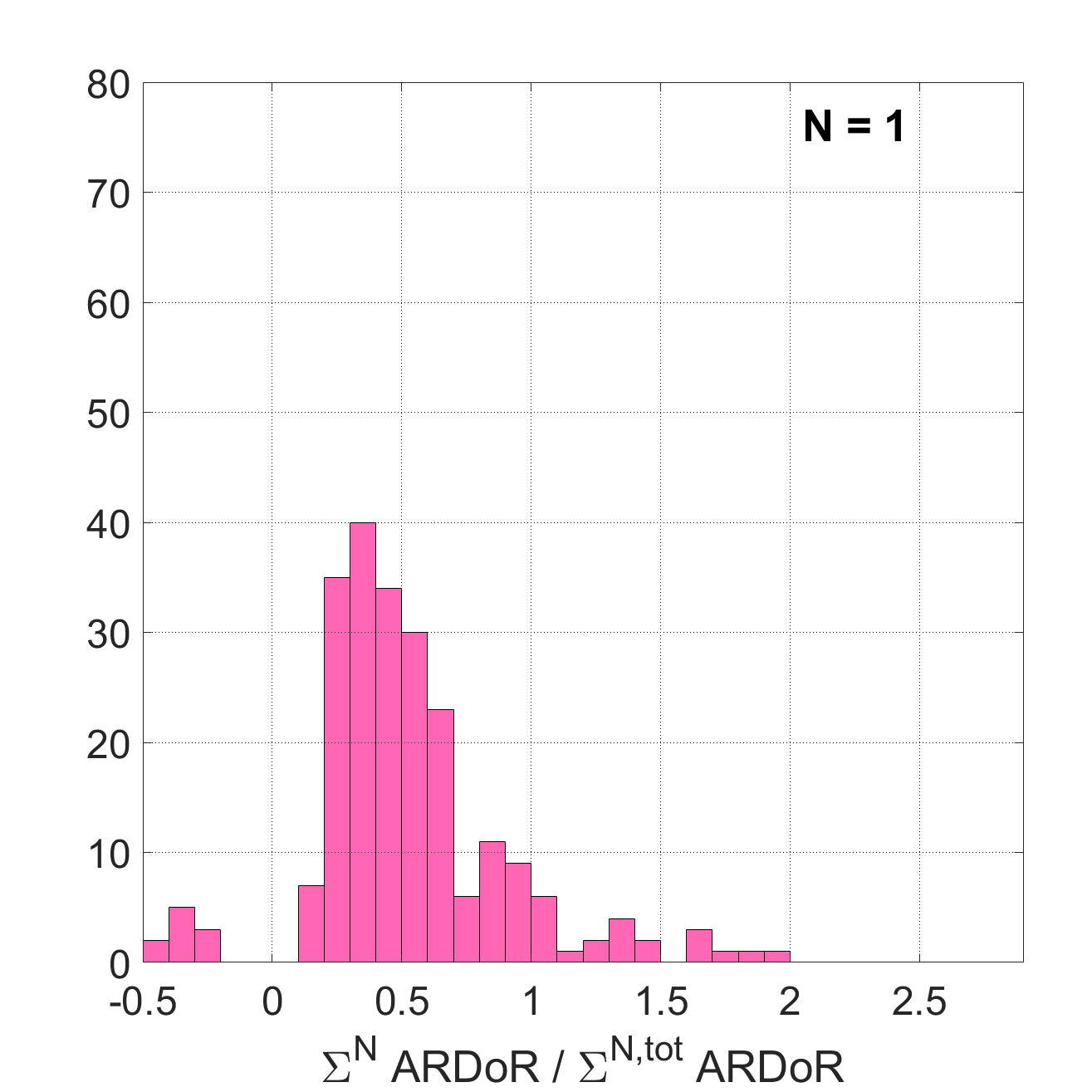}
\includegraphics[width=0.4\textwidth]{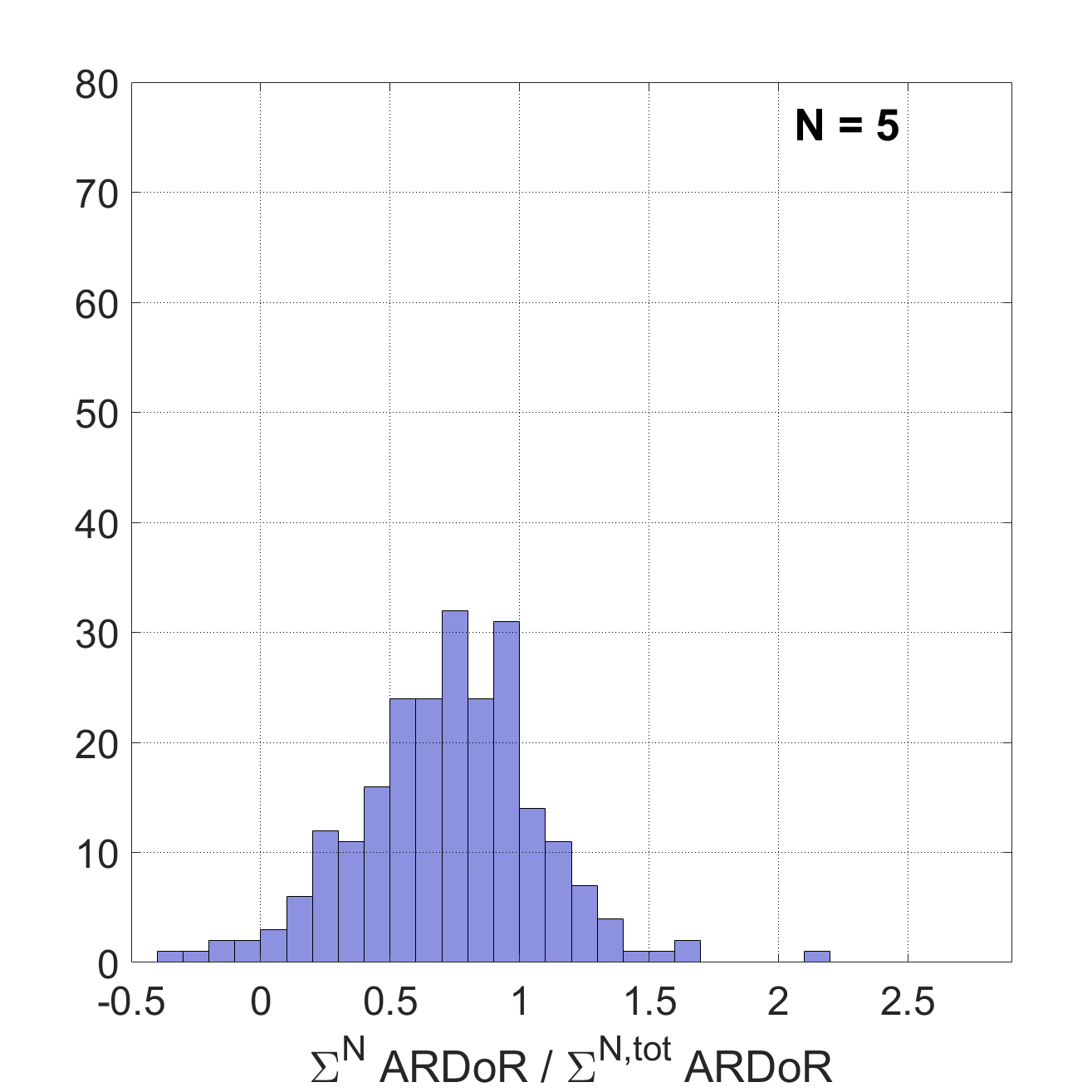}
\includegraphics[width=0.4\textwidth]{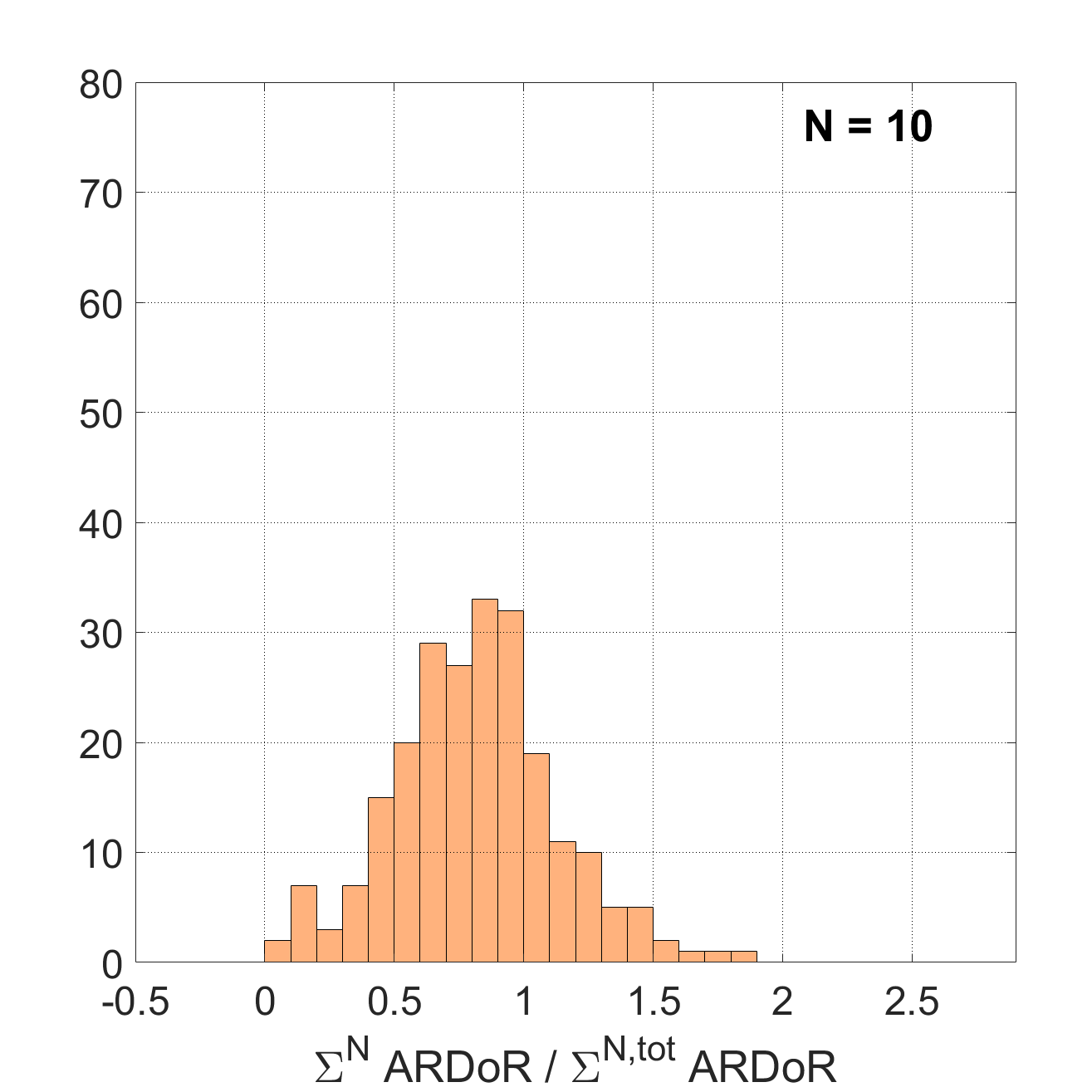}
\includegraphics[width=0.4\textwidth]{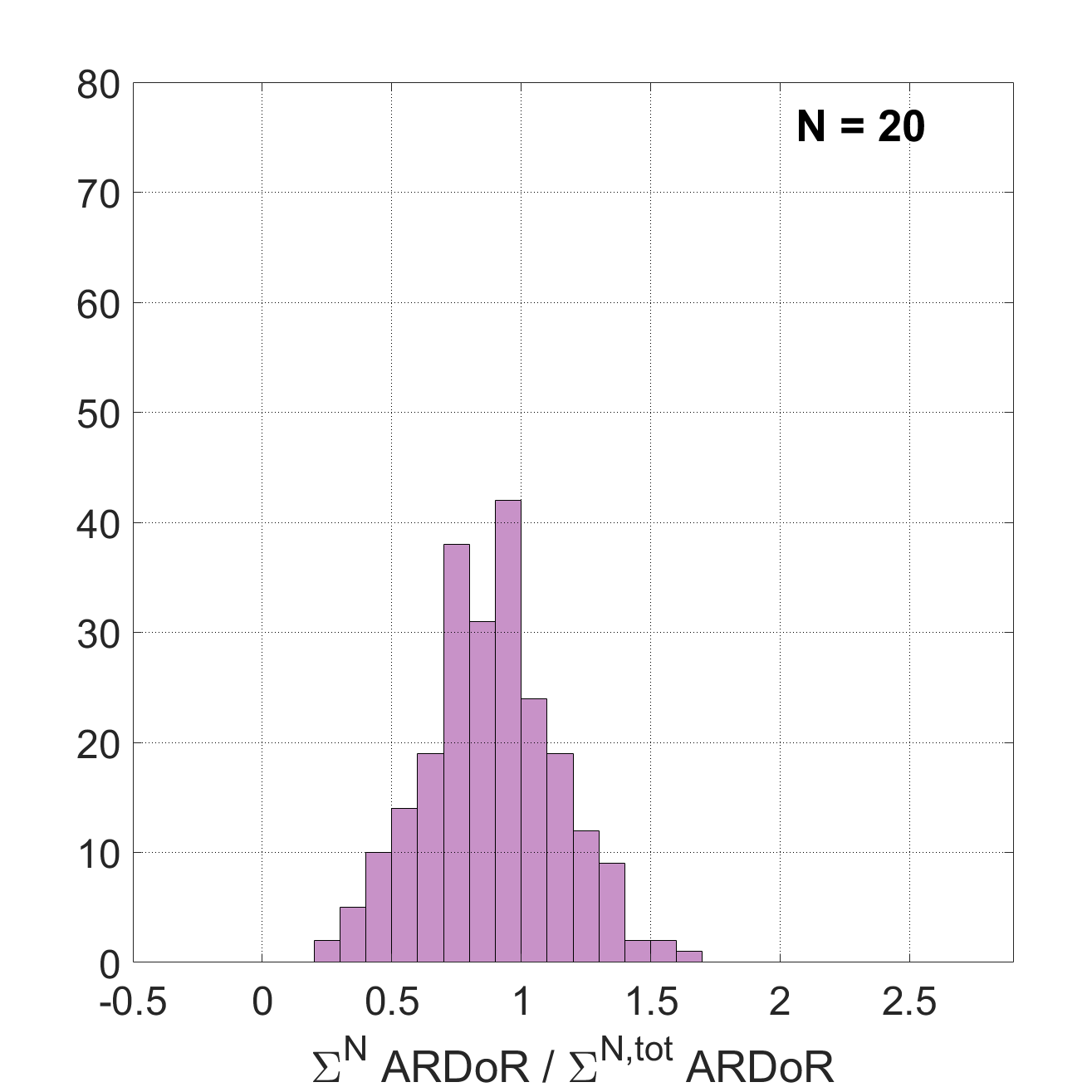}
\includegraphics[width=0.4\textwidth]{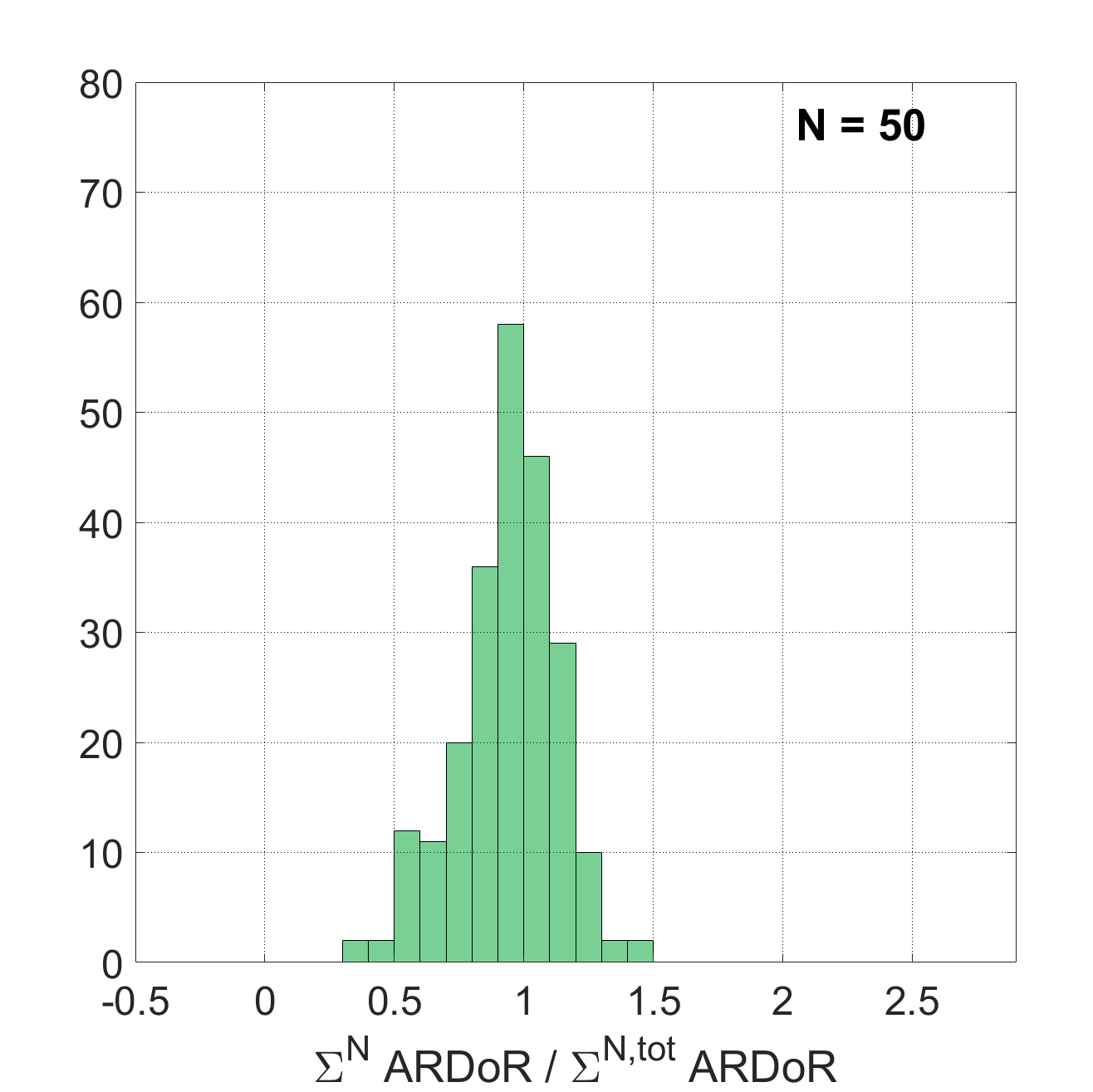}
\includegraphics[width=0.4\textwidth]{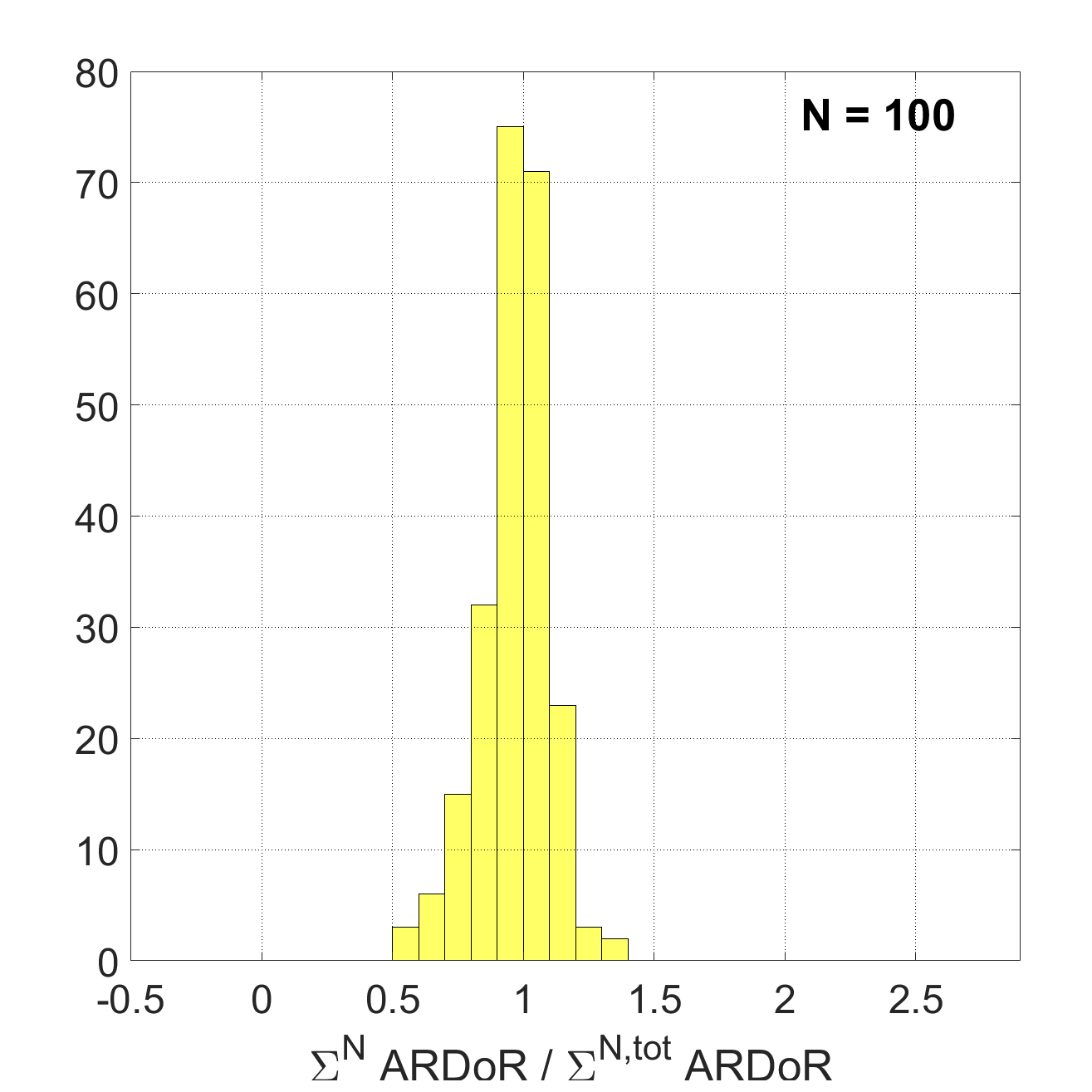}
\end{center}
\caption{Histograms of the total ARDoR of the top $N$ ARs divided by
the total ARDoR of all ARs for cycles where the total ARDoR exceeds
15\,\% of the absolute change in the dipole moment $\Delta D$ (230
cycles). The value of $N$ is shown inside each panel.}
%  Kicsit sok a panel, 2,3,4-et el lehetne hagyni!
\label{fig:ARDORhist15}
\end{figure}

Note that the special case $N=0$, i.e.\ the RS set was already
considered in Paper 1 where we found that even this method yields
$\Delta D$ values in good agreement with the full simulations for a
large majority of the cycles, but the prediction breaks down for a
significant minority. As we are primarily interested in improving
predictions for this minority, we first select cycles where the
difference between the $\Delta D$ \fix{values from the fully stochastic %(FS)
and reduced stochasticity} %(RS) 
sets exceeds $\pm 15\,$\%. (Note that
this difference is by definition equal to the sum of $\ardor$s for all
ARs in the cycle, so the condition for selection was 
$(\sum_{i=1}^{N_\mathrm{tot}}\mathrm{ARDoR}_i)/\Delta D > 0.15$, 
which held for 230 cycles.)

Figure~\ref{fig:ARDORhist15} presents histograms of the fraction of
the deviation explained by ARs with the the top $N$ highest ARDoR
values.  Means, medians and standard deviations of these plots are
collected in Table~\ref{table:ARDORhist15}. It is apparent that ARs
with the top 10--20 highest $\ardor$ are sufficient to explain
80--90\,\% of the deviation of $\Delta D$ computed with the reduced
stochasticity model from the full value of $\Delta D$. Even the single
AR with the highest $\ardor$ alone can explain 50\,\% of the
deviation. Meanwhile, a significant scatter is present in the plots:
{e.g., adding up the columns below 0.5 and above 1.5 in the
4th panel one finds that} 
in $\sim 8\,$\% of these 230 deviating cycles even the rank-20
$\ardor$ method is insufficient to reproduce the deviation at an
accuracy better than $\pm 50$\,\%. (This is $\pm 50$\% of the 
{\it deviation}: as
in this sample the mean deviation is roughly $\sim20$\% of the
expected value of $\Delta D$,  $\Delta D$ itself is still
reproduced with an accuracy up to $\pm 10$\% for these cycles.)

\begin{figure}
\includegraphics[width=\textwidth]{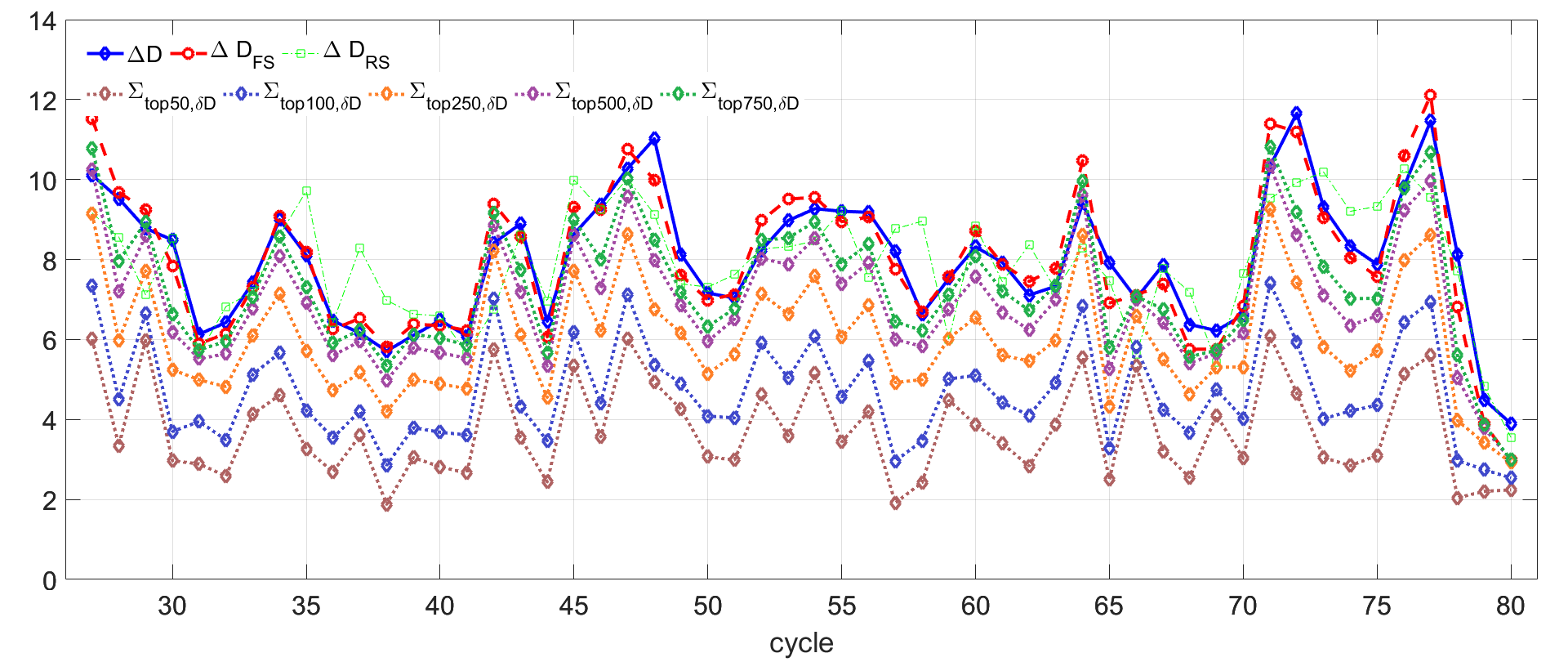}
\includegraphics[width=\textwidth]{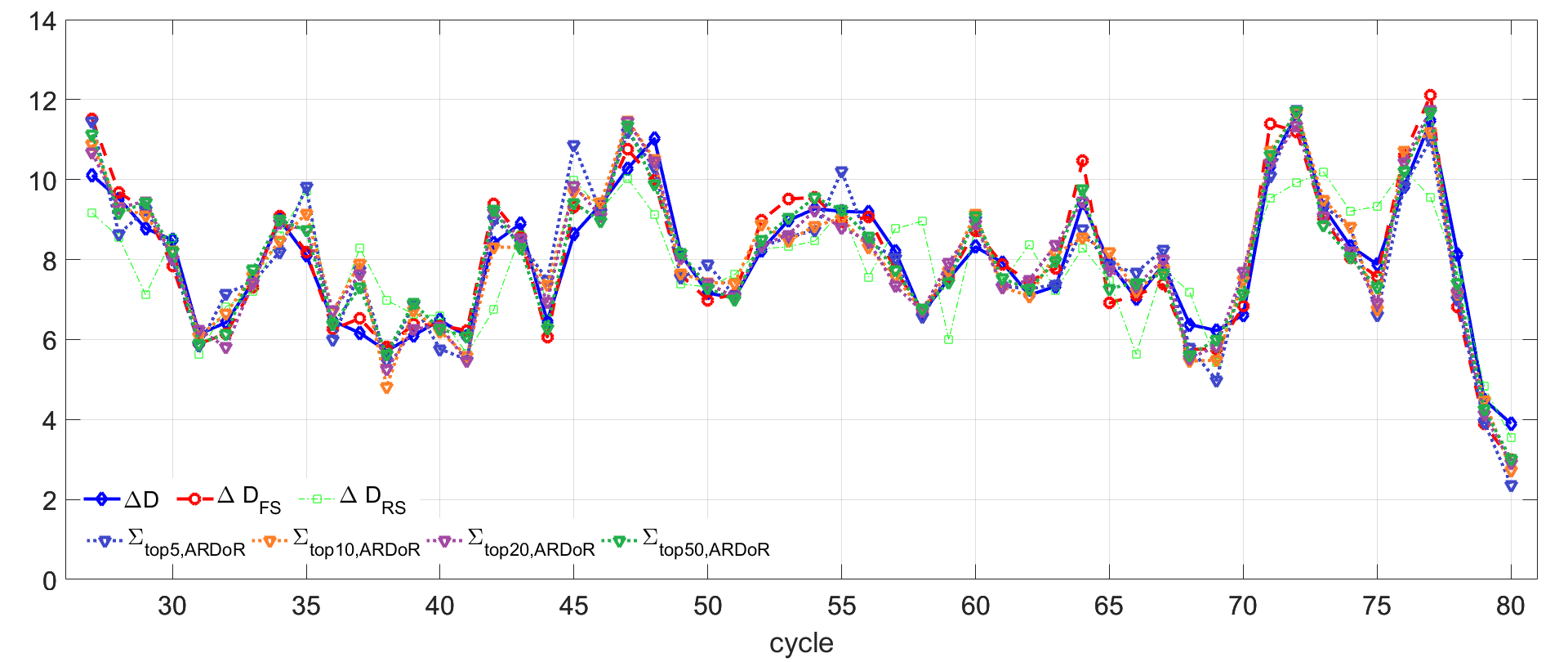}
\caption{Absolute change $\Delta D$ on the global dipole moment during
a cycle (blue solid); its approximation using the fully stochastic
algebraic method (red dashed) and the reduced stochasticity algebraic
method (light green dashed). The curves are compared with the absolute
change computed with the algebraic method for various sets of the
active regions in a cycle.
{\it Top panel:} subsets containing ARs with the top $N$ highest ultimate
dipole contribution (FS case).
{\it Bottom panel:} \fix{subsets containing ARs} with the top $N$
highest $ARDoR$ values added \fix{to $\Delta D_{\mathrm{RS}}$}.}
\label{fig:DDcurves}
\end{figure}

\begin{table}
\centering
\begin{tabular}{lrrrc} 
\multicolumn{5}{l}{$\Delta D_N= \Delta D_\mathrm{RS}+\sum_{i=1}^N\ardor_i$, 
ranking by $\ardor$:}\\
\hline
 $N$ & mean & median & st.dev. & st.dev. of $(\Delta D_N-\Delta
 D)/\Delta D$\\
\hline
  0  &   0.0578&    0.0966 & 1.1365 & 0.212 \\
  5  & --0.0115 &   0.0073 & 0.7360 & 0.128 \\
 10  & --0.0330 & --0.0116 & 0.7134 & 0.120 \\
 20  & --0.0368 & --0.0144 & 0.6662 & 0.125 \\
 50  & --0.0599 & --0.0406 & 0.6132 & 0.110 \\
$ N_\mathrm{tot}$ & --0.0499 & --0.1817 & 0.5861 & 0.101 \\
\hline
&&&\\
\multicolumn{4}{l}{$\sum_{i=1}^N \delta D_{U,i}$,
ranking by $\delta_U$:}\\
\hline
 $N$ & mean & median & st.dev.\\
\hline
  50 & 4.1280 & 4.1777 & 1.4684 \\
 100 & 3.1874 & 3.1173 & 1.3479 \\
 250 & 1.8958 & 1.7491 & 1.1311 \\
 500 & 1.0129 & 0.8471 & 0.9346 \\
 750 & 0.6003 & 0.4895 & 0.8186 \\
\hline\\
\end{tabular}
\caption{Means, medians and standard deviations of the residuals of
various approximations relative to the simulated value of the absolute
dipole moment change $\Delta D$ during an activity cycle, as plotted
in Fig.~\ref{fig:DDcurves}.}
\label{table:DDcurves}
\end{table}

The improvement that the $\ardor$ method brings to the problem of
reproducing the solar axial dipole moment at the end of a cycle is
dramatically illustrated in Fig.~\ref{fig:DDcurves}. While in the case
of ranking ARs by $\delta D$ even adding contributions from the top
750 AR yields only a barely tolerable representation of the dipole
moment variation, the $\ardor$ method produces excellent agreement
already for very low values of the rank $N$. The quality of these
representations is documented in Table~\ref{table:DDcurves}. The
standard deviation of the rank-5 $\ardor$ method relative to the
simulation result is lower than in the case of $\Delta D$ calculated
from even the top 750 highest $\delta D$ contributions.

\begin{figure}
\includegraphics[width=\textwidth]{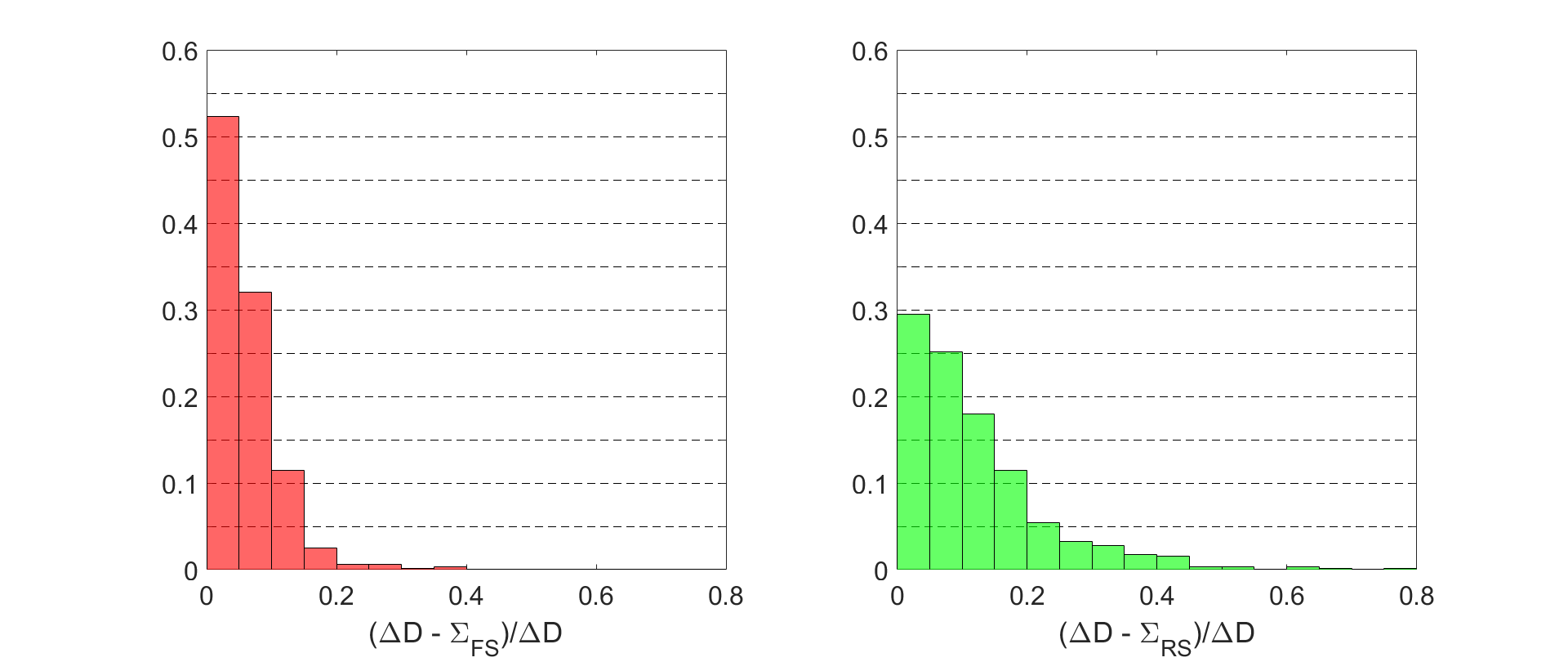}
\includegraphics[width=\textwidth]{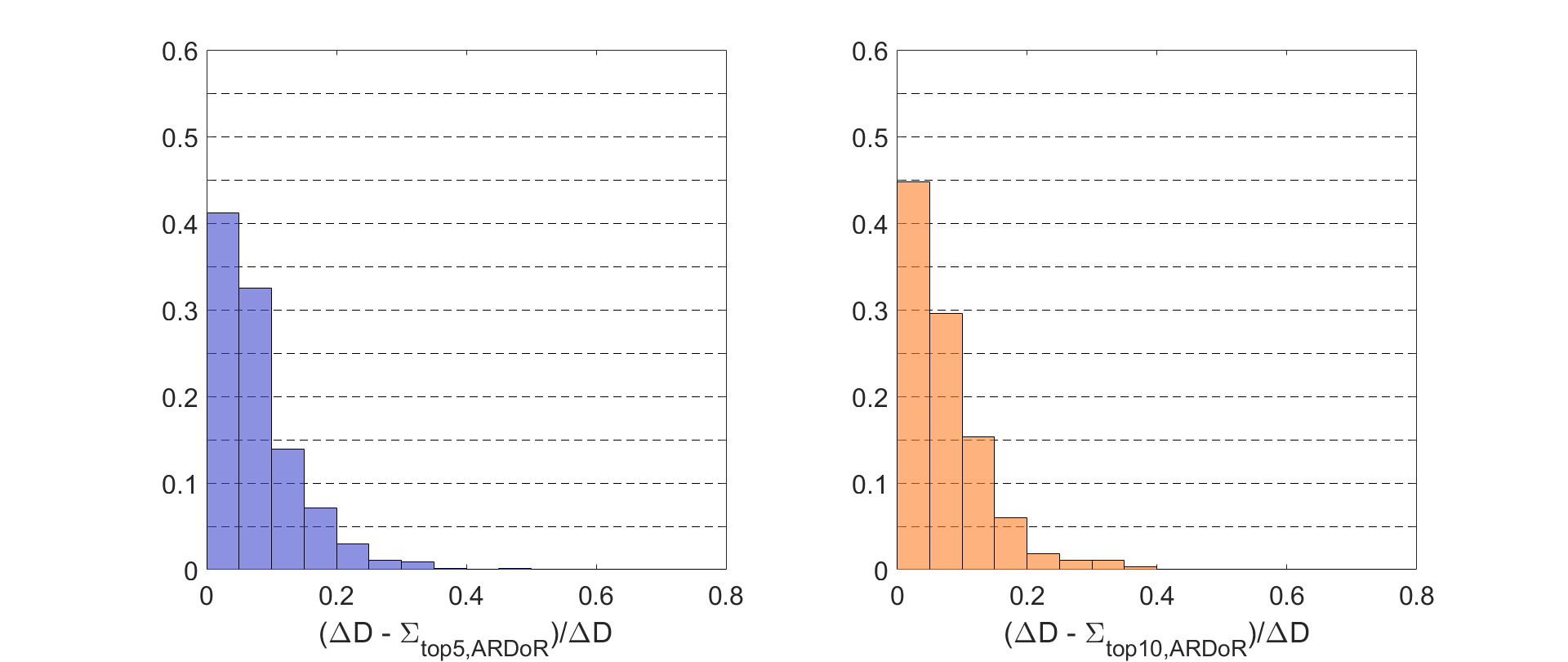}
\includegraphics[width=\textwidth]{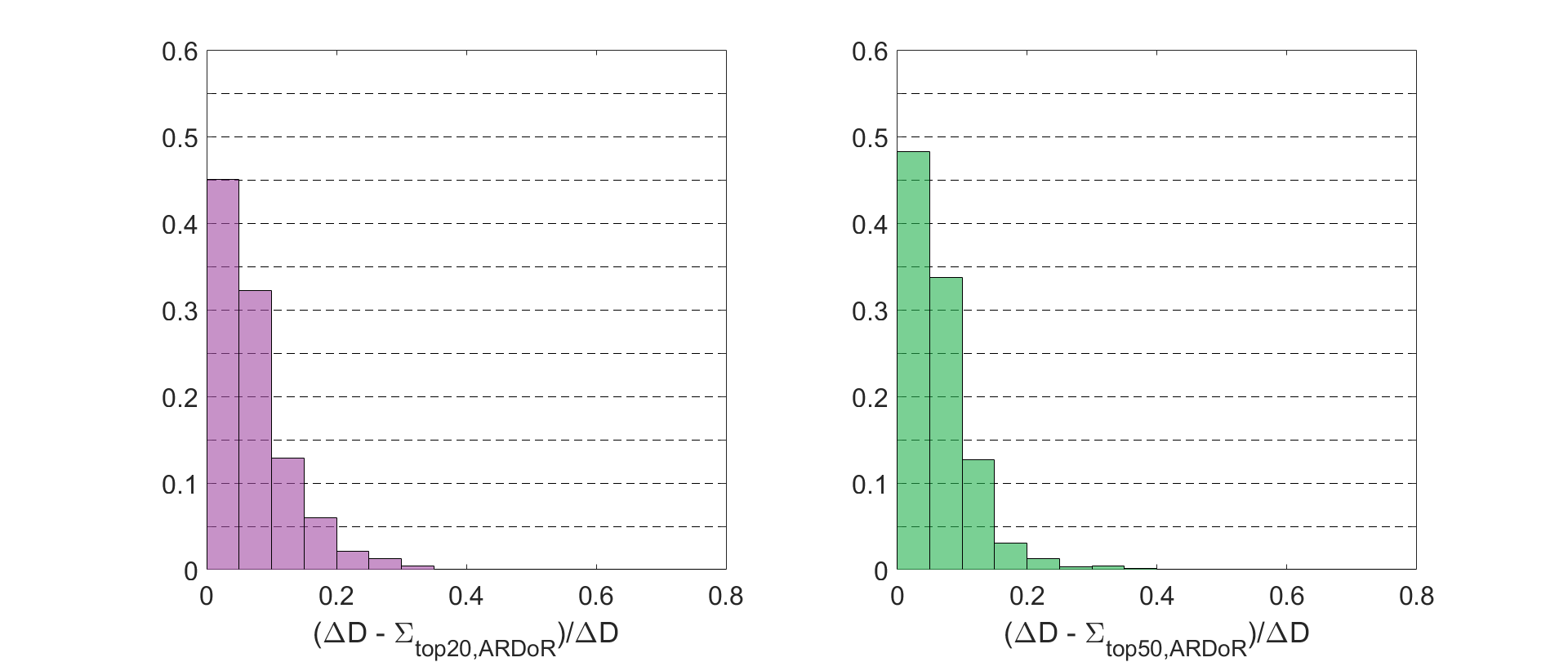}
\caption{Fractional histograms of the fractional deviation from the
absolute change in dipole moment during a solar cycle, calculated
summing ultimate AR contributions in the RS case $+$ ARDoR values for
ARs with the the top $N$ highest ARDoR. 
{($N=N_\mathrm{tot}$ and $N=0$ for the first and second panels, 
respectively.)} 
Colour codes are the same as
in the previous plots.}
\label{fig:fracthist}
\end{figure}

Finally, in Fig.~\ref{fig:fracthist} we present histograms of the
deviations from the simulated value of $\Delta D$ computed with the
various methods (FS, RS and ARDoR with different $N$ values). Here
deviations are expressed as fractions of the actual $\Delta D$
resulting from the simulations, i.e. the quantities given in the
headings of Table~\ref{table:DDcurves} are divided by $\Delta D$.
Adding up the columns it is straightforward to work out from this
that, e.g., in the case of 
{the rank-5 ARDoR method (i.e.,} 
considering only the top 5 highest ARDoR
values and adding them to the RS algebraic result), the deviation of
$\Delta D$ from the simulated cycle change in the global dipole moment
is
less than 15\,\% in 88\,\% of the cycles. As $\Delta D$ is, on
average, twice the amplitude of the polar field at minimum, the rank-5
$\ardor$ method reproduces the  polar field precursor {within
$\pm 30$\,\%} in 88\,\% of all cycles. This is to be compared to 74\,\%
{ of the cycles} in the RS case.

\section{Conclusions}

In Paper 1 we introduced a method to reconstruct variations in the
global axial dipole moment of the Sun by an algebraic summation of the
contributions from individual active regions.  In principle, for the
application of this method, for each AR the optimal representation in
terms of a simple bipole (or possibly several bipoles in more complex
cases) needs to be known. Obtaining this information for thousands of
active regions is a nontrivial task, but significant efforts have
been made in this direction:

\begin{list}{--}{\itemsep=0 em \parsep=0em \partopsep=0em \topsep=0em}
\item \cite{Sheeley+:BMRcyc21} determined the properties of bipoles
representing {close to 3000} ARs with $\Phi>3\cdot
10^{20}\,$Mx from NSO-KP (Kitt Peak) magnetograms in Cycle 21
(1976--1986). Each AR was considered at its maximum development;
recurrent ARs were multiply listed.

\item \cite{Yeates+:BMRcyc23_24} determined the properties of bipoles
representing ARs from NSO-KP/SOLIS synoptic magnetic maps  in cycles
23 and 24 (1997--2017). Each AR was considered at central meridian
passage; recurrent ARs were multiply listed. 

\item \cite{Whitbread+:dipcontr} determined initial dipole moments $D_1$
for active regions from Kitt Peak/SOLIS synoptic magnetic maps in
cycles 21--24 (1976--2017). Each AR was considered at central meridian
passage; recurrent ARs were multiply listed.

\item From white-light data without direct magnetic information
\cite{Jiang+Baranyi} determined an indicative ``dipole moment index''
for sunspot groups larger than 800 MSH in cycles 21--24
(1976--2017). 
\end{list}

Data resulting from the above listed efforts have been placed in 
public databases.\footnote{VizieR and the Solar dynamo dataverse 
({\tt https://dataverse.harvard.edu/dataverse/solardynamo}), 
maintained by Andr\'es Mu\~noz-Jaramillo.} 
In addition to these, \cite{Li+Ulrich:tilts} determined tilt angles
for 30,600 ARs from  Mt.Wilson ad MDI magnetograms in cycles 21--24
(1974--2010). \cite{Virtanen+:dipmom} determined initial dipole
moments $D_1$ for active regions from Kitt Peak/SOLIS synoptic
magnetic maps combined with SDO HMI synoptic maps in cycles 21--24
(1976--2019). 

The above studies are limited to the last four cycles when
magnetograms were available on a regular basis.  For earlier cycles, a
number of statistical analyses of sunspot data without direct magnetic
information (e.g., \citealt{Dasi-Espuig+}, \citealt{Ivanov:noTQ},
\citealt{McClintock+Norton:MtWtilts}, \citealt{Baranyi:tilts},
\citealt{Isik:tilts}, \citealt{SenthamizhPavai}) resulted in tilt
angle values, offering some potential for use as input for models of
the dipole moment evolution. Recently, information on the magnetic
polarities of sunspots from Mt.Wilson measurements has been used in
combination with Ca II spectroheliograms by \cite{Pevtsov+:pseudmgrams}
to construct ``pseudo-magnetograms'' for the period 1915--1985; the
results have been benchmarked against direct observations for the last
period (\citealt{Virtanen+:test}).

Despite these impressive efforts, the determination of AR dipole
moment values to be used as input in our algebraic method is subject
to many uncertainties. As discussed in the Introduction, the available
data are increasingly incomplete for smaller ARs. The arbitrariness of
the time chosen for the incorporation of ARs is also problematic as
during their evolution the structure of ARs can change significantly
due to processes not represented in the SFT models (flux emergence or
localized photospheric flows). The complexities of AR structure imply
that their representation with a \fix{single bipole may be subject to
doubt  (cf.\ \citealt{Jiang+Baranyi}, \citealt{Iijima+:asym})}. And for historical data these
difficulties are further aggravated.

In view of these considerable difficulties, looking for ways to
minimize the need for detailed input data for our algebraic method is
advisable. With this objective in mind, in the present work we
introduced the $\ardor$ method and tested it on a large number of
activity cycles simulated with the 2$\times$2D dynamo model. We found
that 

\begin{list}{--}{\itemsep=0 em \parsep=0em \partopsep=0em \topsep=0em}
\item Including all information on the bipolar active regions appearing
in a cycle, our algebraic method can reproduce the dipole moment at the
end of the cycle with an error below $\pm 30$\,\% in over 97\,\% of cycles.

\item Using only positions and magnetic fluxes of the ARs, and
arbitrarily equating their polarity separations and tilts to their
expected values (reduced stochasticity {or RS} case), the
algebraic method can reproduce the dipole moment at the end of the
cycle with an error below $\pm 30$\,\% in about 74\,\% of cycles.

\item Combining the RS case with detailed information on a small number
$N$ of ARs with the largest ARDoR values, the fraction of unexplained
cycles is significantly reduced (from 26\,\% to
12\,\% in the case of $N=5$ and a {$\pm 30$}\,\% accuracy threshold).
\end{list}

These results indicate that stochastic effects on the intercycle
variations of solar activity are dominated by the effect of a low
number of large ``rogue'' active regions, rather than the combined
effect of numerous small ARs. 

% "horizon of reproducibility"?

Beyond the academic interest of these results, the method has a
potential for use in solar cycle prediction. For the realization of
this potential, however, a number of further problems need to be
addressed. As in forecasts the positions and fluxes of ARs are also
not known, the representation of the majority of ARs not faithfully
represented in our method must be stochastic also in these variables,
or simply replaced by a smooth continuous distribution. Furthermore,
for the selection of ARs with the top $N$ ARDoR values these values
should be theoretically be computed for all ARs. To avoid this need,
``proxies'' of ARDoR based on straightforward numerical criteria may
need to be identified to select the ARs for which a more in-depth
study is then needed to determine ARDoR values. Studies in this
direction are left for further research.

\begin{acknowledgements}

This research was supported by the Hungarian National Research,
Development and Innovation Fund (grant no. NKFI K-128384) and by the
European Union's Horizon 2020 research and innovation programme under
grant agreement No. 739500. The collaboration of the
authors was facilitated by support from the International Space
Science Institute in ISSI Team 474. 

\end{acknowledgements}

%%    This version assumes use of bibtex with the swsc.bib file being present
%%    If your bib file has a different name you need to change the following line

%\bibliography{algebraic2}

%\end{linenumbers}

\end{document}